\documentclass[12pt, a4paper]{article}

\begin{document}

\title{\bf{Coherent states  \\\`a la Klauder-Perelomov  \\for the 
P\"oschl-Teller potentials}}
\author{A. H. El Kinani$^{*}$ and M. Daoud$^{\dagger }$\vspace{0.5cm} \\
$^{*}$L.P.T, Physics Department, Faculty of Sciences, \\
University Mohammed V, P.O.Box 1014\\
Rabat, Morocco.\vspace{1cm}\\
$^{\dagger }$L.P.M.C, Physics Department, Faculty of Sciences, \\
University Ibn Zohr, P.O.Box 28/S\\
Agadir, Morocco.}
\date{}
\maketitle

\begin{abstract}
In this paper we present a scheme for constructing the coherent states of
Klauder-Perelomov's type for a particle which is trapped in P\"oschl-Teller
potentials.
\end{abstract}

\section{Introduction}

Coherent states are one of the important concepts in physics today (for
review see references $\left[ 1,2,3\right] $). The original coherent states
based on the Heisenberg-Weyl group has been extended for a number of Lie
groups with square integrable representations, and they have many
applications in quantum mechanics. In particular, they are used as bases of
coherent states path integrals $\left[ 4\right] $ or dynamical wavepackets
for describing the quantum systems in semiclassical approximations $\left[
5\right] $.

Many definitions of coherent states exist. The first one defines the usual
coherent states as the eigenstates of the annihilation operator $a^{-}$ for
each individual oscillator mode of the electromagnetic field

\begin{equation}
a^{-}\left| z\right\rangle =z\left| z\right\rangle
\end{equation}
Here $\left[ a^{-},a^{+}\right] =1$ ($\left( a^{-}\right) ^{\dagger 
}=a^{+}$%
) and $z$ is a complex constant with conjugate $\bar z$. The unit normalized
states $\left| z\right\rangle $ are given by
\begin{equation}
\left| z\right\rangle =e^{-\frac{\left| z\right| ^2}2}\sum\limits_{n=0}^%
\infty \frac{z^n}{\sqrt{n!}}\left| n\right\rangle
\end{equation}
where $\left| n\right\rangle $ is an element of the Fock space $\mathcal{H}%
\equiv \left\{ \left| n\right\rangle ,n\geq 0\right\} $.

A second definition of coherent states for oscillators assumes the existence
of a unitary $^{^{\prime \prime }}$displacement$^{^{\prime \prime }}$
operator

\begin{equation}
D\left( z\right) =\exp \left( za^{+}-\bar za^{-}\right)
\end{equation}
whose action on the ground states $\left| 0\right\rangle $ give
the coherent states $\left( 2\right) $ parametrized by $z$. The
unitarity of $D(z)$ ensures the correct normalization of $\left|
z\right\rangle $. In view of the canonical commutation relation
$\left[ a^{-},a^{+}\right] =1$, the second definition is
equivalent to the first one.

A third definition is based on the Heisenberg uncertainty relation with the
position $x$ and momentum $p$ given, as usual, by

\begin{equation}
x=\frac 1{\sqrt{2}}\left( a^{-}+a^{+}\right) \hspace{0.6cm}\hbox{and}%
\hspace{0.6cm}p=\frac i{\sqrt{2}}\left( a^{+}-a^{-}\right)
\hbox{,}
\end{equation}
the coherent states defined above have the minimum-uncertainty
value $\Delta x\Delta p=\frac 12$ and maintain this relation in
time (temporal stability of coherent states). Coherent states have
two important properties. First, they are not orthogonal to each
other. Second, they provide a resolution of the identity, i.e.,
they form an overcomplete set of states. Following the first
approach, a formalism, in which coherent states are defined as
eigenstates of a lowering operator, has been proposed by Gazeau
and Klauder for an arbitrary quantum mechanical system $\left[
6\right] $ (see also the references $\left[ 7-9\right] $).

Recently, we gave a complete classification of states minimizing
the Robert \\-son-Schr\"odinger uncertainty relation for an
exactly solvable quantum system $\left[ 10-11\right] $. We
obtained the so-called $^{^{\prime \prime }}$the generalized
intelligent states$^{^{\prime \prime }}$. This approach follows
the definition of coherent states related to the optimization of
Heisenberg relation of familiar harmonic oscillator coherent
states. But, up to now, a formalism defining the coherent states
for an arbitrary quantum system as the action of some displacement
operator on a reference state (coherent states of
Klauder-Perelomov kind) has not been considered in the literature
as far as we know.

Hence, the main purpose of this letter is the construction of the coherent
states of Klauder-Perelomov's type for a quantum mechanical system evolving
in the P\"oschl-Teller potentials.

The paper is organized as follows. In section 2, we review the factorization
of the P\"oschl-Teller Hamiltonian and we introduce creation and
annihilation operators which will be useful for our purpose. Section 3 is
devoted to the definition of the set of coherent states of
Klauder-Perelomov's Type for the quantum system under consideration. In
section 4, we discuss and compare the three definitions of coherent states.
Other properties such as stability in time, overcompleteness, and resolution
to the unity are discussed. Concluding remarks are summarized in the last
section.

\section{Creation and annihilation operators}

We start by recalling the eigenvalues and eigenvectors of a particle trapped
in the P\"oschl-Teller potentials of trigonometric type. The corresponding
Hamiltonian is given by $\left[ 12\right] $

\begin{equation}
H=-\frac{d^2}{dx^2}+V_{\kappa ,\kappa ^{^{\prime }}}(x)
\end{equation}
where $V_{\kappa ,\kappa ^{^{\prime }}}(x)$ is the family of P\"oschl-Teller
potentials indexed continuously by the parameters $\kappa >1$ and $\kappa
^{^{\prime }}>1:$%
\begin{equation}
V_{\kappa ,\kappa ^{^{\prime }}}(x)=\left\{
\begin{array}{c}
\frac 1{4a^2}\left[ \frac{\kappa \left( \kappa -1\right) }{\sin ^2\left(
\frac x{2a}\right) }+\frac{\kappa ^{^{\prime }}(\kappa ^{^{\prime }}-1)}{%
\cos ^2\left( \frac x{2a}\right) }\right] -\frac{(\kappa +\kappa ^{^{\prime
}})^2}{4a^2}\hspace{0.3cm},\hspace{0.3cm}0<x<\pi a \\
\infty \hspace{5cm}x\leq 0\hspace{0.5cm},\hspace{0.5cm}x\geq \pi a
\end{array}
\right.
\end{equation}
It is also called, sometimes, the P\"oschl-Teller of first type.
Clearly, this a smooth approximation for $\kappa ,$ $\kappa
^{^{\prime }}\rightarrow 1^{+}$ of the infinite square-well
potentials over the interval $\left[ 0,\pi a\right] .$ The
P\"oschl-Teller potentials is closely related to several other
potentials which are widely used in molecular and solid state
physics like (i) The symmetric P\"oschl-Teller potentials well
$(\kappa =\kappa ^{^{\prime }}\geq 1)$, (ii) The Scarf potentials
$\frac 12\leq \kappa ^{^{\prime }}<1$ $\left[ 13\right] ,$ (iii)
The modified P\"oschl-Teller potentials which can be obtained by
replacing the trigonometric functions by their hyperbolic
counterparts $\left[ 12,14\right] $, (iv) The Rosen-Morse
potential which is the symmetric modified P\"oschl-Teller
potentials $\left[ 15\right] $. Many interesting properties of the
P\"oschl-Teller potentials was recently reexamined at classical as
well as at quantum levels, in a nice paper by J-P Antoine et al
$\left[ 8\right] .$ The Hamiltonian $H$ can be written in the
following factorized form
\begin{equation}
H=a_{\kappa ,\kappa ^{^{\prime }}}^{+}a_{\kappa ,\kappa ^{^{\prime }}}^{-}
\end{equation}
where the annihilation $a_{\kappa ,\kappa ^{^{\prime }}}^{-}$ and the
creation $a_{\kappa ,\kappa ^{^{\prime }}}^{+}$ operators are given by

\begin{equation}
a_{\kappa ,\kappa ^{^{\prime }}}^{\pm }=\mp \frac d{dx}+W_{\kappa ,\kappa
^{^{\prime }}}(x)
\end{equation}
in terms of the superpotentials $W_{\kappa ,\kappa ^{^{\prime }}}(x)$

\begin{equation}
W_{\kappa ,\kappa ^{^{\prime }}}(x)=\frac 1{2a}\left[ \kappa \hbox{cotg}%
\left( \frac x{2a}\right) -\kappa ^{^{\prime }}\hbox{tang}\left(
\frac x{2a}\right) \right]
\end{equation}
The normalized eigenstates $\psi _n(x)$ and the corresponding eigenvalues $%
e_n$ are given by

\begin{equation}
\psi _n(x)=\left[ c_n(\kappa ,\kappa ^{^{\prime }})\right] ^{-\frac
12}\left[ \cos \left( \frac x{2a}\right) \right] ^{\kappa ^{^{\prime
}}}\left[ \sin \left( \frac x{2a}\right) \right] ^\kappa P_n^{(\kappa -\frac
12,\kappa ^{^{\prime }}-\frac 12)}\left( \cos \left( \frac xa\right) \right)
\end{equation}
where $c_n(\kappa ,\kappa ^{^{\prime }})$ is a normalization factor that can
be evaluated, and
\begin{equation}
e_n=n\left( n+\kappa +\kappa ^{^{\prime }}\right) \hspace{1cm}n=0,1,2,...
\end{equation}
In Eq. $\left( 10\right) $, the $P_n^{(\alpha ,\beta )}$'s stand
for the Jacobi polynomials.

We define the actions of the creation and annihilation operators on the
eigenstates $\left| \psi _n\right\rangle $ as follows
\begin{eqnarray}
a_{\kappa ,\kappa ^{^{\prime }}}^{+}\left| \psi _n\right\rangle &=&\sqrt{%
\left( n+1\right) \left( n+1+\kappa +\kappa ^{^{\prime }}\right) }%
e^{-i\alpha (2n+1+\kappa +\kappa ^{^{\prime }})}\left| \psi
_{n+1}\right\rangle \\
a_{\kappa ,\kappa ^{^{\prime }}}^{-}\left| \psi _n\right\rangle &=&\sqrt{%
n\left( n+\kappa +\kappa ^{^{\prime }}\right) }e^{i\alpha (2n-1+\kappa
+\kappa ^{^{\prime }})}\left| \psi _{n-1}\right\rangle
\end{eqnarray}
where the real parameter $\alpha $ plays an important role in the temporal
stability of coherent states. This remark will be clarified in the sequel of
this letter. From the equations $\left( 12\right) $ and $\left( 13\right) $,
one can verify that the operators $a_{\kappa ,\kappa ^{^{\prime }}}^{+}$ and
$a_{\kappa ,\kappa ^{^{\prime }}}^{-}$satisfy the following commutation
relation
\begin{equation}
\left[ a_{\kappa ,\kappa ^{^{\prime }}}^{-}a_{\kappa ,\kappa ^{^{\prime
}}}^{+}\right] =G_{\kappa ,\kappa ^{^{\prime }}}\left( N\right)
\end{equation}
where
\begin{equation}
G_{\kappa ,\kappa ^{^{\prime }}}\left( N\right) \equiv G=2N+\kappa +\kappa
^{^{\prime }}+1
\end{equation}
and the operator $N$ is defined by his action on states $\left| \psi
_n\right\rangle $ as
\begin{equation}
N\left| \psi _n\right\rangle =n\left| \psi _n\right\rangle
\end{equation}
We note that $N\neq a_{\kappa ,\kappa ^{^{\prime }}}^{+}a_{\kappa ,\kappa
^{^{\prime }}}^{-}=H.$

\section{Coherent states of Klauder-Perelomov's type}

In view of the second definition of coherent states (CS) for the standard
harmonic oscillator, we define the CS of Klauder-Perelomov's type for a
particle submitted to P\"oschl-Teller potentials as follows
\begin{equation}
\left| z,\alpha \right\rangle =\exp \left( za_{\kappa ,\kappa ^{^{\prime
}}}^{+}-\overline{z}a_{\kappa ,\kappa ^{^{\prime }}}^{-}\right) \left| \psi
_0\right\rangle \hspace{1cm},\hspace{0.5cm}z\in \mathbf{C}
\end{equation}
So, we have to compute the action of the unitary operator
\begin{equation}
D(z)=\exp \left( za_{\kappa ,\kappa ^{^{\prime }}}^{+}-\overline{z}a_{\kappa
,\kappa ^{^{\prime }}}^{-}\right)
\end{equation}
on the ground state $\left| \psi _0\right\rangle $ of the system
under consideration. Using the action of the annihilation and the
creation operators on the Hilbert space $\mathcal{H}=\left\{
\left| \psi _n\right\rangle ,n=0,1,2,...\right\} $ (Eqs. $\left(
12\right) $ and $\left( 13\right) $), one can show that the state
$\left| z,\alpha \right\rangle $ can be written, in a compact
form, as
\begin{equation}
\left| z,\alpha \right\rangle =\sum\limits_{n=0}^{+\infty
}a_n\left( \left| z\right| \right) z^ne^{-i\alpha e_n}\left| \psi
_n\right\rangle
\end{equation}
The quantities $a_n\left( \left| z\right| \right) $ in $\left( 19\right) $
are defined by
\begin{equation}
a_n\left( \left| z\right| \right) =\sqrt{\frac{\Gamma \left( n+1\right)
\Gamma \left( n+\kappa +\kappa ^{^{\prime }}+1\right) }{\Gamma \left( \kappa
+\kappa ^{^{\prime }}+1\right) }}c_n\left( \left| z\right| \right)
\end{equation}
and the coefficients $c_n\left( \left| z\right| \right) $ are given by
\begin{equation}
c_n\left( \left| z\right| \right) =\sum\limits_{j=0}^{+\infty
}\frac{\left( -\left| z\right| ^2\right) ^j}{\left( n+2j\right)
!}\left(
\sum\limits_{i_1=1}^{n+1}e_{i_1}\sum\limits_{i_2=1}^{i_1+1}e_{i_2}....%
\sum\limits_{i_j=1}^{i_{j-1}+1}e_{i_j}\right)
\end{equation}
Setting
\begin{equation}
\pi
(n+1,j)=\sum\limits_{i_1=1}^{n+1}e_{i_1}\sum%
\limits_{i_2=1}^{i_1+1}e_{i_2}....\sum\limits_{i_j=1}^{i_{j-1}+1}e_{i_j}%
\hspace{0.3cm},\hbox{and }\hspace{0.3cm}\pi \left( n+1,0\right) =1
\end{equation}
one can verify that the $\pi $'s satisfy the following relation
\begin{equation}
\pi \left( n+1,j\right) -\pi \left( n,j\right) =(n+1)(n+\kappa +\kappa
^{^{\prime }}+1)\pi \left( n+2,j-1\right)
\end{equation}
Using this recurrence formula, it is not difficult to show that the $%
c_n\left( \left| z\right| \right) ^{\prime }$s satisfy the following
differential equation
\begin{equation}
\left| z\right| \frac{dc_n\left( \left| z\right| \right) }{d\left| z\right| 
}%
=c_{n-1}\left( \left| z\right| \right) -nc_n\left( \left| z\right| \right)
-(n+1)(n+1+\kappa +\kappa ^{^{\prime }})\left| z\right| ^2c_{n+1}\left(
\left| z\right| \right)
\end{equation}

Hence, solving this equation, we can explicitly obtain the P\"oschl-Teller
coherent states of Klauder-Perelomov's type. The solutions of the
differential equation $\left( 24\right) $ for $\kappa +\kappa ^{\prime }\in
\mathbf{N}\setminus \left\{ 0,1,2\right\} $, are of the form
\begin{equation}
c_n(\left| z\right| )=\frac 1{n!\left| z\right| ^n}\ss _{m,n+\frac 12(\kappa
+\kappa ^{^{\prime }}+1)}^{-\frac 12(\kappa +\kappa ^{^{\prime }}+1)}\left(
\cosh (2\left| z\right| )\right) \hspace{0.4cm},
\end{equation}
because the Jacobi functions $\mathbf{\ss }$ satisfy
\begin{equation}
\frac d{d\left| z\right| }\ss _{m,n-l}^l(\cosh \left( 2\left|
z\right| \right) )=n\ss _{m,n-1-l}^l(\cosh \left( 2\left| z\right|
\right) )-\left( n+2l\right) \ss _{m,n+1-l}^l(\cosh \left( 2\left|
z\right| \right) )
\end{equation}
where $l=-\frac 12(\kappa +\kappa ^{^{\prime }}+1)$ and $m$ is
free integer
or half-integer parameter. It is clear that the differential equation $%
\left( 24\right) $ admits several solutions depending on $m$. However, an
admissible solution in our case is obtained by noting that 
$D(z=0)=\mathbf{1}
$ (unit operator). Recall that the Jacobi functions are defined by $\left[
16\right] $%
\begin{eqnarray}
\ss _{m,n}^l(\cosh \left( 2x\right) ) &=&\Gamma \left(
l+n+1\right) \Gamma \left( l-n+1\right) \left( \cosh x\right)
^{2l}\left( \tanh x\right) ^{n-m}\times  \\ &&\ \ \ \ \
\sum\limits_{s=\max (0,m-n)}^{+\infty }\frac{\left( \tanh x\right)
^{2s}\left( \Gamma \left( s+1\right) \right) ^{-1}\left( \Gamma
\left( l-n-s+1\right) \right) ^{-1}}{\Gamma \left( n-m+s+1\right)
\Gamma \left( l+m+1\right) }\hbox{,}  \nonumber
\end{eqnarray}
where $l$ is any complex number, and $m,n$ are both integer or both
half-integers. These functions play an important role in the representation
theory for the group $QU(2)$ of unimodular quasi-unitary matrices. Then,
using the definition $\left( 27\right) $, we obtain the unique solution
compatible with the condition $D(z=0)=\mathbf{1}$%
\begin{equation}
c_n(\left| z\right| )=\frac 1{n!\left| z\right| ^n}\ss _{\frac 12(\kappa
+\kappa ^{^{\prime }}+1),n+\frac 12(\kappa +\kappa ^{^{\prime }}+1)}^{-\frac
12(\kappa +\kappa ^{^{\prime }}+1)}\left( \cosh (2\left| z\right| )\right)
\end{equation}
which can be written also as
\begin{equation}
c_n(\left| z\right| )=\frac 1{n!}\left( \cosh (\left| z\right| )\right)
^{-(\kappa +\kappa ^{^{\prime }}+1)}\left( \frac{\tanh \left| z\right| }{%
\left| z\right| }\right) ^n
\end{equation}
The coherent states of Klauder-Perelomov's type take now the form
\begin{eqnarray}
\left| z,\alpha \right\rangle  &=&\left( 1-\tanh ^2\left| z\right|
\right) ^{\frac 12(\kappa +\kappa ^{^{\prime
}}+1)}\sum_{n=0}^{+\infty }\left( \frac{z\tanh \left| z\right|
}{\left| z\right| }\right) ^n\times   \nonumber
\\
&&\ \ \ \ \left[ \frac{\Gamma (n+1+\kappa +\kappa ^{^{\prime }})}{\Gamma
\left( n+1\right) \Gamma \left( 1+\kappa +\kappa ^{^{\prime }}\right) }%
\right] ^{\frac 12}e^{-i\alpha n(n+\kappa +\kappa ^{^{\prime }})}\left| \psi
_n\right\rangle
\end{eqnarray}
Finally, setting $\mathbf{\zeta }=\frac{z\tanh \left| z\right| }{\left|
z\right| }$, we obtain
\begin{eqnarray}
\left| \mathbf{\zeta },\alpha \right\rangle  &\equiv &\left| z,\alpha
\right\rangle =\left( 1-\left| \mathbf{\zeta }\right| ^2\right) ^{\frac
12(\kappa +\kappa ^{^{\prime }}+1)}\sum\limits_{n=0}^{+\infty }\mathbf{%
\zeta }^n\left[ \frac{\Gamma (n+1+\kappa +\kappa ^{^{\prime }})}{\Gamma
\left( n+1\right) \Gamma \left( 1+\kappa +\kappa ^{^{\prime }}\right) }%
\right] ^{\frac 12}\times   \nonumber \\ &&\ \hbox{ }e^{-i\alpha
n(n+\kappa +\kappa ^{^{\prime }})}\left| \psi _n\right\rangle
\end{eqnarray}
The parameter $\mathbf{\zeta }$ is restricted by $\left| \mathbf{\zeta }%
\right| <1.$

The identity resolution is
\begin{equation}
\int \left| \mathbf{\zeta },\alpha \right\rangle \left\langle
\mathbf{\zeta },\alpha \right| d\mu \left( \mathbf{\zeta }\right)
=I_{\mathcal{H}}
\end{equation}
where the measure is given by
\begin{equation}
d\mu \left( \mathbf{\zeta }\right) =\frac{\kappa +\kappa ^{^{\prime }}}\pi
\frac{d^2\mathbf{\zeta }}{\left( 1-\left| \mathbf{\zeta }\right| ^2\right) 
^2%
}
\end{equation}
There are two main consequence arising from the former result. First, one
can express any coherent state $\mid \mathbf{\zeta }^{^{\prime }},\alpha
^{^{\prime }}\rangle $ in terms of the others
\begin{equation}
\mid \mathbf{\zeta }^{^{\prime }},\alpha ^{^{\prime }}\rangle
=\int \left| \mathbf{\zeta },\alpha \right\rangle \left\langle
\mathbf{\zeta },\alpha \right. \mid \mathbf{\zeta }^{^{\prime
}},\alpha ^{^{\prime }}\rangle d\mu \left( \mathbf{\zeta }\right)
\end{equation}
The kernel $\left\langle \mathbf{\zeta },\alpha \right. \mid \mathbf{\zeta 
}%
^{^{\prime }},\alpha ^{^{\prime }}\rangle $ is easy to evaluate from $\left(
31\right) $%
\begin{eqnarray}
\left\langle \mathbf{\zeta },\alpha \right. &\mid &\mathbf{\zeta }^{^{\prime
}},\alpha ^{^{\prime }}\rangle =\left( 1-\left| \mathbf{\zeta }\right|
^2\right) ^{\frac 12(\kappa +\kappa ^{^{\prime }}+1)}\left( 1-\left| 
\mathbf{%
\zeta }^{^{\prime }}\right| ^2\right) ^{\frac 12(\kappa +\kappa ^{^{\prime
}}+1)}\sum\limits_{n=0}^{+\infty }\left( \overline{\mathbf{\zeta }}\mathbf{%
\zeta }^{^{\prime }}\right) ^n\times  \nonumber \\
&&\ \frac{\Gamma (n+1+\kappa +\kappa ^{^{\prime }})}{\Gamma \left(
n+1\right) \Gamma \left( 1+\kappa +\kappa ^{^{\prime }}\right) }e^{-i(\alpha
+\alpha ^{^{\prime }})n(n+\kappa +\kappa ^{^{\prime }})}
\end{eqnarray}
The coherent states are normalized $\left( \left\langle \mathbf{\zeta }%
,\alpha \right| \left. \mathbf{\zeta },\alpha \right\rangle =1\right) $, but
they are not orthogonal to each other.

Second, an arbitrary element state of the Hilbert space $\mathcal{H}$ , let
us call it $\left| f\right\rangle $, can be written in terms of the coherent
states
\begin{equation}
\left| f\right\rangle =\int f\left( \mathbf{\zeta
},\overline{\mathbf{\zeta
}}\right) \left| \mathbf{\zeta },\alpha \right\rangle d\mu \left( \mathbf{%
\zeta }\right)
\end{equation}
where the analytic function
\begin{eqnarray}
f\left( \mathbf{\zeta },\overline{\mathbf{\zeta }}\right) &=&\left( 1-\left|
\mathbf{\zeta }\right| ^2\right) ^{\frac 12(\kappa +\kappa ^{^{\prime
}}+1)}\sum\limits_{n=0}^{+\infty }\overline{\mathbf{\zeta }}^n\left[ \frac{%
\Gamma (n+1+\kappa +\kappa ^{^{\prime }})}{\Gamma \left( n+1\right) \Gamma
\left( 1+\kappa +\kappa ^{^{\prime }}\right) }\right] ^{\frac 12}\times
\nonumber \\
&&\ e^{-i\alpha n(n+\kappa +\kappa ^{^{\prime }})}\left\langle \psi
_n\right| \left. f\right\rangle
\end{eqnarray}
determines completely the state $\left| f\right\rangle \in $ $\mathcal{H}$ .

Let us now consider the dynamical evolution of the coherent states. More
precisely, we have
\begin{equation}
U\left( t\right) \left| \mathbf{\zeta },\alpha \right\rangle =e^{-iHt}\left|
\mathbf{\zeta },\alpha \right\rangle =\left| \mathbf{\zeta },\alpha
+t\right\rangle
\end{equation}
The coherent states are stable temporally.

\section{Other kinds of coherent states}

It is well known that, and as we mentioned in the introduction, there are
three different group-theoretic approaches to coherent states $\left[
1-3\right] $. These approaches follow three possible definitions of the
familiar Glauber coherent states $\left[ 17\right] $ of a harmonic
oscillator. In a similar way, we have developed (section 2) the formalism in
which coherent states, for P\"oschl-Teller potentials, are generated by the
action of the displacement operator on the ground state. In a second
approach, developed recently by Gazeau and Klauder $\left[ 6-7\right] $, one
deals with eigenstates of the annihilation operator. This second definition
follows the Barut-Girardello construction of the $su(1,1)$ coherent states 
$%
\left[ 18\right] $. The third definition of coherent states is related to
the optimization of Robertson-Schr\"odinger (R-S) uncertainty relation $%
\left[ 19-20\right] $. States that minimize (R-S) uncertainty relation are
called intelligent states $\left[ 10-11,21-24\right] $. These different
definitions of the coherent states for an arbitrary quantum system coincide
only in the special case of the Weyl-Heisenberg algebra that is the
dynamical symmetry of a quantized harmonic oscillator. For other quantum
mechanical systems, like a particle in the P\"oschl-Teller potentials, these
different approaches lead to distinct states. Relation between various sets
of coherent states should be studied by developing a formalism that provides
a unified description of these different states for an arbitrary quantum
system. This matter remains an open problem.

In this section for review purposes, we give the main properties of
Gazeau-Klauder coherent states $\left[ 6\right] $ and intelligent states
(states minimizing the R-S uncertainty relation) $\left[ 10\right] $, for
the P\"oschl-Teller potentials in order to compare various sets of coherent
states obtained following the standard three definitions.

\subsection{Gazeau-Klauder coherent states}

The so-called Gazeau-Klauder coherent states $\left[ 6-7\right] $, defined
as eigenstates of the annihilation operator $a_{\kappa ,\kappa ^{^{\prime
}}}^{-}$

\begin{equation}
a_{\kappa ,\kappa ^{^{\prime }}}^{-}\left| z,\alpha \right\rangle =z\left|
z,\alpha \right\rangle \hspace{1cm}z\in \mathbf{C,}\hspace{0.5cm}\alpha \in
\mathbf{R}
\end{equation}
are given by $\left[ 8,\hbox{see also }10\right] $%
\begin{equation}
\left| z,\alpha \right\rangle =\mathcal{N}\left( \left| z\right|
\right) \sum\limits_{n=0}^{+\infty }\frac{z^ne^{-i\alpha
n(n+\kappa +\kappa ^{^{\prime }})}}{\sqrt{\Gamma \left( n+1\right)
\Gamma \left( n+\kappa +\kappa ^{^{\prime }}+1\right) }}\left|
\psi _n\right\rangle
\end{equation}
for a particle trapped in the P\"oschl-Teller potentials of trigonometric
type ($\kappa ,\kappa ^{^{\prime }}>1$). The normalization constant $%
\mathcal{N}\left( \left| z\right| \right) $ is given by
\begin{equation}
\left[ \mathcal{N}\left( \left| z\right| \right) \right] ^2=\frac{\left|
z\right| ^{\kappa +\kappa ^{^{\prime }}}}{I_{\kappa +\kappa ^{^{\prime
}}}\left( 2\left| z\right| \right) }
\end{equation}
where $I_{\kappa +\kappa ^{^{\prime }}}\left( 2\left| z\right| \right) $ is
the modified Bessel function of the first kind.

The resolution of unity is explicitly given by
\begin{equation}
\int \left| z,\alpha \right\rangle \left\langle z,\alpha \right|
d\mu \left( z\right) =I_{\mathcal{H}}
\end{equation}
where the measure can be computed by using the inverse Mellin transform $%
\left[ 25\right] $ (for more details, see $\left[ 10-11\right] $).
\begin{equation}
d\mu \left( z\right) =\frac 2\pi I_{\kappa +\kappa ^{^{\prime }}}\left(
2r\right) K_{\frac{\kappa +\kappa ^{^{\prime }}}2}\left( 2r\right) rdrd\phi 
\hspace{1cm};\hspace{0.5cm}z=re^{i\phi }
\end{equation}
The Gazeau-Klauder coherent states are continuous in the labelling $z$ and 
$%
\alpha $. They form an overcomplete family of states (identity resolution)
and satisfy the identity action
\begin{equation}
\left\langle z,\alpha \right| H\left| z,\alpha \right\rangle =\left|
z\right| ^2
\end{equation}
One can verify that the states $\left| z,\alpha \right\rangle $ (eq $40$)
are stable temporally, i.e.
\begin{equation}
e^{-iHt}\left| z,\alpha \right\rangle =\left| z,\alpha +t\right\rangle
\end{equation}
Clearly, the Gazeau-Klauder coherent states $\left( 40\right) $ are
different from the ones constructed \`a la Klauder-Perelomov $\left(
31\right) $.

\subsection{States minimizing the Robertson-Schr\"odinger uncertainty
relation}

Using the creation and annihilation ($a_{\kappa ,\kappa ^{^{\prime }}}^{+}$
and $a_{\kappa ,\kappa ^{^{\prime }}}^{-}$) operators, we introduce two
hermitian operators
\begin{equation}
W\equiv W_{\kappa ,\kappa ^{^{\prime }}}=\frac 1{\sqrt{2}}\left( a_{\kappa
,\kappa ^{^{\prime }}}^{+}+a_{\kappa ,\kappa ^{^{\prime }}}^{-}\right) %
\hspace{1.5cm}P\equiv P_{\kappa ,\kappa ^{^{\prime }}}=\frac i{\sqrt{2}%
}\left( a_{\kappa ,\kappa ^{^{\prime }}}^{+}-a_{\kappa ,\kappa ^{^{\prime
}}}^{-}\right)
\end{equation}
which satisfy the commutation relation
\begin{equation}
\left[ W,P\right] =iG_{\kappa ,\kappa ^{^{\prime }}}\equiv iG
\end{equation}
It is well known that the variances $\left( \Delta W\right) ^2$ and $\left(
\Delta P\right) ^2$ obey the Robertson-Schr\"odinger uncertainty relation
\begin{equation}
\left( \Delta W\right) ^2\left( \Delta P\right) ^2\geq \frac 14\left(
\left\langle G\right\rangle ^2+\left\langle F\right\rangle ^2\right)
\end{equation}
where the operator $F$ is defined by
\begin{equation}
F=\left\{ W-\left\langle W\right\rangle ,P-\left\langle P\right\rangle
\right\}
\end{equation}
The symbol $\left\{ ,\right\} $ stands for the anticommutator (for more
details see the references $\left[ 10-11\right] $). The states minimizing
the Robertson-Schr\"odinger uncertainty relation satisfy the eigenvalues
equation
\begin{equation}
\left( W+i\lambda P\right) \left| z,\lambda ,\alpha \right\rangle 
=z\sqrt{2}%
\left| z,\lambda ,\alpha \right\rangle
\end{equation}
which can be written also as
\begin{equation}
\left[ \left( 1-\lambda \right) a_{\kappa ,\kappa ^{^{\prime }}}^{+}+\left(
1+\lambda \right) a_{\kappa ,\kappa ^{^{\prime }}}^{-}\right] \left|
z,\lambda ,\alpha \right\rangle =2z\left| z,\lambda ,\alpha \right\rangle
\end{equation}
The parameter $\lambda \in \mathbf{C}$ is called, some-times, the squeezing
parameter.

In the states satisfying $\left( 50\right) $, we have the following
relations
\begin{equation}
\left( \Delta W\right) ^2=\left| \lambda \right| \Delta \hspace{1.5cm}\left(
\Delta P\right) ^2=\frac 1{\left| \lambda \right| }\Delta
\end{equation}
with
\begin{equation}
\Delta =\frac 12\sqrt{\left\langle G\right\rangle ^2+\left\langle
F\right\rangle ^2}
\end{equation}
A complete classification of the solutions (generalized intelligent states)
of $\left( 50\right) $ was obtained in $\left[ 10\right] $ for an arbitrary
exactly solvable quantum system. Here, we are interested by the situation
where $\left| \lambda \right| =1$. In this case, we have
\begin{equation}
\left( \Delta W\right) ^2=\left( \Delta P\right) ^2
\end{equation}
and the states satisfying the equation $\left( 50\right) $ with $\left|
\lambda \right| =1$ are called generalized coherent states. They are given
by $\left[ 10\right] $ (up to the normalization constant)
\begin{equation}
\left| z,\lambda ,\alpha \right\rangle =U\left( z,\lambda \right) \left|
\psi _0\right\rangle
\end{equation}
where the operator $U\left( z,\lambda \right) $, providing the state $\left|
z,\lambda ,\alpha \right\rangle $ by acting on the ground state $\left| \psi
_0\right\rangle $, is given by
\begin{equation}
U\left( z,\lambda \right) =\sum_{n=0}^{+\infty }\left[ \left( \frac{2z}{%
1+\lambda }\right) \frac 1Ha_{\kappa ,\kappa ^{^{\prime }}}^{+}+\left( 
\frac{%
\lambda -1}{\lambda +1}\right) \frac 1H\left( a_{\kappa ,\kappa ^{^{\prime
}}}^{+}\right) ^2\right] ^n
\end{equation}
Coherence and squeezing of the generalized intelligent states was discussed
in $\left[ 10\right] $. Here also it is clear that the generalized coherent
states ($\left| \lambda \right| =1$) obtained by minimizing the
Robertson-Schr\"odinger uncertainty relation are different from two sets of
coherent states discussed before (Klauder-Perelomov and Gazeau-Klauder 
ones).

The generalized coherent states $\left| z,\lambda ,\alpha
\right\rangle $ equation $\left( 55\right) $, obtained by
minimizing the Robertson-Schr\"odinger uncertainty relation,
generalize the Gazeau-Klauder ones. Indeed, the latters are
obtained by simply setting $\lambda =1$. The states $\left|
z,\lambda =1,\alpha \right\rangle \equiv $ Eq. $\left( 51\right) $
are eigenstates of the annihilation operator $a_{\kappa ,\kappa
^{^{\prime }}}^{-}$. In this case ($\lambda =1$), we get
\begin{equation}
\left( \Delta W\right) ^2=\left( \Delta P\right) ^2=\frac 12\left\langle
G\right\rangle
\end{equation}
where the average value of $G$ is
\begin{eqnarray}
\left\langle G\right\rangle &=&\left\langle z,\lambda =1,\alpha \right|
G_{\kappa ,\kappa ^{^{\prime }}}\left( N\right) \left| z,\lambda =1,\alpha
\right\rangle  \nonumber \\
\ &=&\left( 1+\kappa +\kappa ^{^{\prime }}\right) +\frac{2\left| z\right| 
^2%
}{\left( 1+\kappa +\kappa ^{^{\prime }}\right) }\frac{_0F_1\left( 2+\kappa
+\kappa ^{^{\prime }},\left| z\right| ^2\right) }{_0F_1\left( 1+\kappa
+\kappa ^{^{\prime }},\left| z\right| ^2\right) }
\end{eqnarray}
in terms of the confluent hypergeometric function $_0F_1\left( a,x\right) $ 
$%
\left[ 16\right] .$

Note that
\begin{equation}
\left\langle G\right\rangle \geq 1+\kappa +\kappa ^{^{\prime }}
\end{equation}
which traduce the fact that the dispersions $\left( \Delta W\right) ^2$ and 
$%
\left( \Delta P\right) ^2$ are greater than $\frac 12$, unlike the
case of the harmonic oscillator in which we have $\left( \Delta
W\right) ^2=\left( \Delta P\right) ^2=\frac 12.$

\section{Concluding remarks}

In this article, we have defined coherent states of Klauder-Perelomov's type
of a particle trapped in P\"oschl-Teller potentials. They present an
important differences from the ones derived \`a la Gazeau-Klauder and by
minimization of the Robertson-Schr\"odinger uncertainty relation. To close
this letter, let us mention that the results obtained here can be extended
to introduce coherent states of Klauder-Perelomov's type for the infinite
square well potential. We should also quote Nieto and Simmons $\left[
26\right] ,$ who have considered the P\"oschl-Teller potentials as exemples
of their construction of coherent states for confining one dimentional
potentials. More recently, the classical limit of coherent states for
P\"oschl-Teller potentials was investigated $\left[ 27\right] $. The states
constructed in $\left[ 26,27\right] $ have a totally different meaning from
the ones discussed in this work. In fact, here we deal with coherent states
of Klauder-Perelomov's kind. Moreover, the so-called Gazeau-Klauder coherent
states and ones minimizing the Robertson-Schr\"odinger uncertainty relation
are more general than ones constructed in $\left[ 26,27\right] .$ We believe
that the recent results on the coherent states constructions $\left[
6-7,10-11\right] $ (see also $\left[ 28\right] $) can be used to understand
the quantum properties as well as the classical limit of the P\"oschl-Teller
coherent states. On other hand, in view of the results discussed through
this work, a natural question arises. For an arbitrary quantum system, what
formalism unifies the description of the various kinds of coherent states
within a commun frame? We believe that the answer to this will establish
relations between different types of coherent states to understand the
physical basis of their mathematical properties. Our work is a first step in
this sense. More details and further developments on this subject will be
submitted for publication in the near future $\left[ 29\right] .$

\begin{quote}
$\mathbf{Acknowledgements}$

M. Daoud would like to thank Professors, J-P Gazeau, V. Hussin and M. Kibler
for valuable discussions. The authors are very grateful to Professor J-P
Antoine for sending the reference $\left[ 8\right] $. They are also indebted
to the referee for interesting comments and suggestions.

\newpage\
\end{quote}


\begin{thebibliography}{99}
\bibitem{a}  J. R. Klauder and B. S. Skagerstam, Coherent
States-Applications in Physics and Mathematical Physics (World Scientific,
Singapore, 1985).

\bibitem{b}  A. Perelomov, Generalized Coherent States and Their
Applications (Texts and Monographs in physics, Singapore, Berlin, 1985).

\bibitem{c}  S. T. Ali, J-P. Antoine and J-P. Gazeau, Coherent States,
Wavelets and Their Generalizations (Springer-Verlag, New York, 2000).

\bibitem{d}  J. R. Klauder, Phys. Rev. D \textbf{19} (1979) 2349.

\bibitem{e}  R. G. Littlejohn, Phys. Rep. \textbf{138} (1986) 193, and
references therein.

\bibitem{f}  J-P. Gazeau and J. R. Klauder, J. Phys. A: Math. Gen. 
\textbf{32%
} (1999) 123.

\bibitem{g}  J. R. Klauder, J. Phys. A: Math. Gen. \textbf{29} (1996) L 293.

\bibitem{h}  J-P. Antoine, J-P. Gazeau, P. M. Monceau, J. R. Klauder and K.
A. Penson, Temporally stable coherent states for infinite well and
P\"oschl-Teller potentials, UCL-IPT-00-03 preprint.

\bibitem{i}  J-P. Gazeau and B. Champagne, '' The Fibonacci-deformed
harmonic oscillator'' in Algebraic Methods in Physics, Y. Saint-Aubin and L.
Vinet (eds.) CRM Series in Theoretical and Mathematical Physics, Vol 
\textbf{%
3} (Springer-Verlag, Berlin, 2000).

\bibitem{j}  A. H. El Kinani and M. Daoud, Generalized intelligent states
for an arbitrary quantum system, L. P .T. preprint 2001, submitted for
publication.

\bibitem{k}  A. H. El Kinani and M. Daoud, Generalized intelligent states
for non-linear oscillators, L. P. T. preprint 2001, submitted for
publication.

\bibitem{l}  G. P\"oschl and E. Teller, Z. Physik \textbf{83} (1933) 143.

\bibitem{m}  F. L. Scarf, Phys. Rev \textbf{112} (1958) 1137.

\bibitem{n}  C. Daskaloyannis, J. Phys. A: Math. Gen. \textbf{25} (1992) 
261.

\bibitem{o}  N. Rosen and Ph. M. Morse, Phys. Rev \textbf{42} (1932) 210.

\bibitem{p}  Vilenkin, N. Ya. Special functions and the theory of group
representations, American Mathematical Society, Providence, Rhode Island
(1968).

\bibitem{q}  R. J. Glauber, Phys. Rev. \textbf{130} (1963) 2529; Phys. Rev.
\textbf{131} (1963) 2766.

\bibitem{r}  A. O. Barut and L. Girardello, Commun. Math. Phys. \textbf{21}
(1971) 41.

\bibitem{s}  E. Schr\"odinger, Sitzungsber. Preuss. Akad. Wiss (Berlin,
1930) p. 296.

\bibitem{t}  H. P. Robertson, Phys. Rev. \textbf{35} (1930) 667; Phys. Rev.
\textbf{46} (1934) 794.

\bibitem{u}  V. V. Dodonov, E. V. Kurmyshev and V. I. Man'ko, Phys. Lett. A
\textbf{79} (1980) 150.

\bibitem{v}  C. Brif, Inter. Jour. Theo. Phys. \textbf{36} (1997) 1651.

\bibitem{w}  D. A. Trifonov, Phys. Lett. A \textbf{48} (1974) 165; J. Math.
Phys. \textbf{35} (1994) 2297.

\bibitem{x}  V. Hussin, Generalized minimum uncertainty relation and a new
class of super-squeezed states, in Proceeding of ''$6^{th}$ International
conference on squeezed states and uncertainty relation, Naples, Italy, May
1999''. To be published.

\bibitem{y}  H. Bateman, Table of integral transforms, Vol \textbf{1}, 1954,
ed A Erd\'elyi (New York, McGraw Hill).

\bibitem{y1}  M. M. Nieto, L. M. Jr. Simmons, Phys. Rev. D \textbf{20}
(1979) 1332.

\bibitem{y3}  M. G. A. Crawford and E. R. Vrscay, Phys. Rev. A \textbf{57}
(1998) 106.

\bibitem{y2}  G. Ghosh, J. Math. Phys. \textbf{39} (1998) 1366.

\bibitem{z}  A. H. El Kinani and M. Daoud, in preparation.
\end{thebibliography}
\end{document}